\documentclass[12pt]{iopart}   

\usepackage{graphicx}
\usepackage{color}
\usepackage{iopams}
\usepackage{bbm}

\begin{document}
\title{On the Nodal Count Statistics for Separable Systems in any Dimension} 
\author{Sven  Gnutzmann and Stylianos Lois} 
\address{School of Mathematical
  Sciences, University of Nottingham, Nottingham NG7 2RD, UK} \date{}

\begin{abstract}
  We consider the statistics of the number of nodal domains aka
  nodal counts
  for eigenfunctions of separable wave equations in arbitrary
  dimension.
  We give an explicit expression for the limiting distribution of
  normalised nodal counts and analyse some of its universal
  properties. Our results are illustrated by detailed discussion
  of simple examples and numerical nodal count distributions.
\end{abstract}
\maketitle

\section{Introduction}

We consider real square-integrable eigenfunctions $\Phi(\mathbf{q})$
of the stationary Schr\"odinger equation
\begin{equation}
  H \Phi(\mathbf{q})\equiv
  - \Delta_{\mathcal{M}}\Phi(\mathbf{q}) + 
  V(\mathbf{q})\Phi(\mathbf{q}) = E \Phi(\mathbf{q}) 
  \label{schrod}
\end{equation}
for a massive point particle on an $s$-dimensional smooth connected
Riemannian manifold $\mathcal{M}$ with local coordinates
$\mathbf{q}\equiv(q^1,\dots,q^s)$ .  Here, $\Delta_{\mathcal{M}}$ is
the Laplace-Beltrami operator on $\mathcal{M}$, $V(\mathbf{q})$ is a
potential, and $E$ is an energy eigenvalue.  We have set the value of
the physical constant $\frac{\hbar^2}{2m}$ of Planck's constant
squared over twice the mass of the particle equal to one by
appropriate choice of units.\\
We will allow that $\mathcal{M}$ has a boundary and will impose
boundary conditions on $\Phi(\mathbf{q})$ such that the Schr\"odinger
operator $H$ defined in (\ref{schrod}) is
self-adjoint (e.g. Dirichlet or Neumann boundary conditions).\\
We consider only non-negative potentials for which the \textit{classically
  allowed region} $\mathcal{V}_E = \left\{ \mathbf{q}:
  V(\mathbf{q})\le E\right\}$ is compact and connected. 
This ensures a discrete and non-negative energy spectrum.  
If $V(\mathbf{q})=0$ (free motion) the
condition implies that the manifold $\mathcal{M}$ is compact.

We arrange the spectrum in ascending order as $0 \le E_1 < E_2 \le
\cdots \le E_N\le E_{N+1}\le \cdots$ and denote by
$\Phi_N(\mathbf{q})$ the eigenfunction corresponding to $E_N$.  For a
given eigenfunction $\Phi_N(\mathbf{q})$ the nodal set $\mathcal{N}=
\Phi_N^{-1}(0)\subset\mathcal{M}$ consists of all points on the
manifold where the eigenfunction vanishes. A nodal domain
$\mathcal{D}\subset \mathcal{M}$ of $\Phi_N(\mathbf{q})$ is a
maximally connected region where the sign of $\Phi_N(\mathbf{q})$ does
not change.

The characterisation of eigenfunctions in terms of their nodal set
has a history which is more than 200 years old with the first
systematic treatment by Chladni \cite{Chladni} who visualised the
vibration modes of plates with sand that accumulates at the nodal set.
Among other things he also counted the number of different nodal
domains for each vibration mode and used these number to characterise
the modes for a given shape.  The number of nodal domains of
eigenfunctions or \emph{nodal counts} will also be the subject of the
present contribution.  For the wave function $\Phi_N$ we denote the
nodal count by $\nu_N$. The collection of all nodal counts forms the
\emph{nodal sequence} $\{ \nu_N \}_{N=1}^\infty$ (for systems with
degenerate eigenvalues this definition of the nodal sequence
is incomplete).\\
In one dimension Sturm's oscillation theorem \cite{Sturm} states
$\nu_N=N$ under very general conditions. The generalisation of this
seminal result to quasi one dimensional systems such as quantum graphs
has been a recent research topic \cite{Berkolaiko, Schapotschikow}.
For arbitrary dimension a seminal result is Courant's nodal domain
theorem \cite{Courant} which states $\nu_N\le N$ for the Laplacian in
any dimension.  Pleijel later showed that in dimension $d= 2$ the
upper bound $\nu_N=N$ is achieved only a finite number of times
\cite{Pleijel}.

In recent years it has been established that the nodal sequence
contains a lot of information about the underlying geometry.
It has been conjectured in \cite{GnSmiSond} that the nodal count sequence
in some cases 
allows a full reconstruction of the manifold $\mathcal{M}$ up to an overall
scaling factor, and that it can be used to distinguish between isospectral
partners. 
These conjectures have been partly confirmed and refined in recent years
\cite{GnuKarSmiPRL,BKP,GnuKarSmi-Wittenberg,KarSmi,Klawonn,BF}.

Another recent line of research focusses on the statistics of the
nodal counts.  To this end one defines \cite{BGS} the \emph{normalised
  nodal count} by the ratio
\begin{equation}
  \xi_N=\frac{\nu_N}{N}
  \label{xi_n}
\end{equation}
and focusses on the distribution of its values in an energy window.
Courant's theorem implies $0<\xi_N\leq 1$.  For a given spectral
interval $I_g(E)=[E,(1+g)E]$ (where $g>0$) one defines the
nodal count distribution formally by
\begin{equation} \label{ndomdistr}
  P(\xi)_{I_g(E)}=\frac{1}{N_{I_g(E)}}\sum_{N: \; E_N \in \; I_g(E)} \delta
  \left(\xi - \xi_N \right)
\end{equation}
where $N_{I_g(E)}$ is the number of eigenvalues in $I_g(E)$. An interesting
question concerns the existence and properties of a (smooth) limiting
distribution
\begin{equation}
  \label{limit_dist}
  P(\xi)=\lim_{E\rightarrow\infty} P(\xi)_{I_g(E)}\ .
\end{equation}
Such a limit may exist (in the weak sense) because the number of
states
in the interval $I_g(E)$ grows as $E \to \infty$.\\
For two-dimensional separable systems a semiclassical theory shows
\cite{BGS, SmiSanka} that the limiting function indeed exists
and that it can be
expressed explicitly in terms of the corresponding
integrable classical dynamics. It has a number of universal features:
\begin{enumerate}
\item The limiting distribution has support $0\le \xi <
  \xi_{\mathrm{crit}}$ where the critical value is smaller than one
  (which is consistent with Pleijel's theorem \cite{Pleijel}).
\item Near the critical value the limiting distribution has a
  square-root singularity
  \begin{equation}
    P(\xi) \sim 
    \left(\xi_{\mathrm{crit}} -\xi \right)^{-1/2} \qquad \mathrm{for}\; \xi<\xi_{\mathrm{crit}}\ .
  \end{equation}
\end{enumerate}
In this work we will generalize this theory to separable
systems in any dimension.

For non-separable systems the semiclassical
theory breaks down -- mainly due
to the lack of an explicit functional that maps
a given eigenfunction to its nodal count.  
In this case one may still find the nodal count numerically
using for instance variants of the Hoshen-Kopelman algorithm
\cite{hoshen}. For two-dimensional
systems with a corresponding classical dynamics that shows chaos
(this is usually referred to as quantum or wave chaos) 
such an approach revealed that the limiting
distribution is universal. Independent of the details of the system
the limiting distribution contracts to a Gaussian located at a
universal value $\xi_{\mathrm{u}}$, i.e. $P(\xi) = \delta(\xi-\xi_u)$
\cite{BGS}.  Consistency with Berry's random wave conjecture
\cite{berry-rw} has also been checked numerically -- the conjecture
states that eigenfunctions of a chaotic billiard follow the same
statistics as the (monochromatic) Gaussian random wave model
(a random superposition of plane waves of the same wavelength).\\
The universality of the nodal count statistics for 
wave-chaotic systems
in two dimensions has been explained in a
seminal work by Bogomolny and Schmit \cite{Bogomolny} who constructed
a heuristic parameter-free critical percolation model that predicts
the numerical value of $\xi_{\mathrm{u}}$ in perfect
agreement with numerical calculations 
and with the Gaussian random wave model (see also
\cite{nodal-keat,foltin,bogomolny2}). Proving rigorously 
the implied conjecture
that the two-dimensional Gaussian random wave model and wave functions
of chaotic billiards are realisations of critical percolation is
certainly one of the most challenging open mathematical questions in the field. 
Indeed a few of the implied properties have already been proven
for random
spherical harmonics \cite{nazarov}.
A related and equally challenging 
conjecture states that the nodal lines for such systems are a
realisation of stochastic Loewner evolution (SLE)
\cite{nodal-keat,bogomolny2,sle-keat,sle-bog,sle-keat2}.
The theoretically known statistical properties of nodal counts 
in two-dimensional wave-chaotic systems have
also been tested thoroughly in experimental settings
\cite{nodal-exp1,nodal-exp2,kuhl}.\\
Preliminary theoretical and numerical results for two-dimensional 
systems that are neither separable nor
nor fully wave chaotic have been obtained for 
non-integrable systems with mixed phase space \cite{aiba} and
for 
integrable systems for which the wave equation is not separable 
\cite{triangle}. Especially the latter shows that
nodal count statistics in non-separable integrable systems 
have a high degree of complexity with a few features that resemble
either the separable or the wave-chaotic case while new features appear.

In this work we address nodal counts in arbitrary dimensions.
Indeed little is known for dimension larger than two.
We will focus on the separable case. 
In Section \ref{sepndom} we derive an asymptotic expression for the
normalised nodal count and related it to the geometry of
the unit energy shell in action space.
In Section \ref{limdist} we will give a general expression for
the limiting nodal count distribution 
and show that it has some
universal properties whose details change with
the dimension.
In Section \ref{examples} the cuboid and the harmonic oscillator are discussed
in more detail and the limiting distribution is compared to numerically
obtained histograms for finite energy intervals.
Eventually
we will discuss in Section \ref{discussion}
some generalisations of our results and also
comment on the nodal
count for wave chaotic systems and 
random waves in higher dimensions.

\section{Nodal domain distributions for separable
  systems}\label{sepndom}

We consider nodal counts for solutions of the wave equation
(\ref{schrod}) in the case where a separation Ansatz leads to the full
solution of the eigenvalue problem.  The tools we will apply for the
derivation of the nodal counts and of the limiting distribution
(\ref{limit_dist}) are EBK quantisation and Poisson summation.  The
asymptotic limit will be an integral over a region in phase space,
and it will involve only classical quantities.  We will 
start with introducing the
relevant classical mechanics.  
We will not try to be as general as possible during the derivation.
Rather we will make some assumptions that simplify the derivation and
later discuss (see Section \ref{discussion}) which assumptions are essential 
and which may be relaxed.

\subsection{EBK quantisation and its implication for nodal counts}

Separability of the wave equation (\ref{schrod}) implies that there exist
coordinates $\mathbf{q}=(q^1,\dots, q^s)$ which (almost) cover the whole
$s$-dimensional manifold $\mathcal{M}$ such that any eigenfunction can
be written in a product form
\begin{equation}
  \Phi(\mathbf{q})=\prod_{l=1}^s\phi^{(l)}(q_l)\ .
\end{equation} 
For such systems semiclassical Einstein-Brillouin-Keller (EBK)
quantisation can be performed successfully. The corresponding
classical Hamiltonian Mechanics on the phase space $T^*\mathcal{M}$
(cotangent bundle to the configuration manifold $\mathcal{M}$) is
generated by the
\begin{equation}
  H(\mathbf{p},\mathbf{q})=\sum_{u,v=1}^s g^{uv}(\mathbf{q})p_u p_v +V(\mathbf{q})
\end{equation}
where $p_l$ is the conjugate momentum to $q^l$, and $g^{uv}$ is the
inverse to the metric tensor $g_{uv}$ which defines the squared
distance $ds^2= \sum_{u,v=1}^s g_{uv}(\mathbf{q})
dq^u dq^v$.\\
Quantum separability implies that the corresponding Hamiltonian
dynamics is integrable. The dynamics
is confined to an $s$-dimensional sub-manifold defined by $s$
independent constants of motion $C_{n}(\mathbf{p},\mathbf{q})= c_n$ in
phase space that (generically) has the topology of a torus. One
introduces the action variables
\begin{equation}
  I_l = \frac{1}{2\pi}\oint_{\mathbf{c}} p_l dq^l
\end{equation}
where the integration is a long the curve in the $p_l$-$q^l$ plane
where it intersects with the torus defined by the values $\mathbf{c}$ for
the constants of motion -- the action is proportional to 
the area enclosed by the torus
in that plane. One may perform a canonical transformation to
action and angle variables $(\mathbf{p},\mathbf{q})\mapsto
(\mathbf{I},\boldsymbol{\theta})$ where
$\boldsymbol{\theta}=(\theta^1,\dots,\theta^s)$ are conjugate to the
actions $\mathbf{I}=(I_1,\dots,I_s)$. I.e. the phase space is foliated
in tori such that a point in phase space is specified by the torus
with action variables $\mathbf{I}$ and the position on the torus
specified in terms of the $s$ angles $\boldsymbol{\theta}$.  As the
action variables are constants of motion all angle variable become cyclic
variables for the transformed Hamilton function $H(\mathbf{I})$.

We will make the following additional assumptions on the classical
Hamiltonian dynamics:
\begin{itemize}
\item[(A1)] The potential is non-negative and the classically allowed
  region $\mathcal{V}_E = \left\{ \mathbf{q} \in \mathcal{M}:
    V(\mathbf{q})\le E\right\}$ is connected and compact. We have stated this
  assumption in the introduction.  This assumption ensures we have a
  discrete non-negative spectrum.
\item[(A2)] There is a one-to-one correspondence between tori in phase
  space and points $\mathbf{I}$ in action space.  This assumption
  excludes double-well potentials and similar potentials in higher
  dimensions where action variables can only be defined locally in
  regions bounded by stationary points and separatrices.
\item[(A3)] A related assumption is the Hamiltonian is a strictly
  increasing function of all action variables
  \begin{equation}
    \omega^l(\mathbf{I})\equiv \frac{\partial H}{\partial I_l} >0 \ .
  \end{equation}
  Here $\omega^l(\mathbf{I})$ is the angular velocity of the angle
  variable $\theta^l$ on the torus defined by $\mathbf{I}$.\\
  We also assume that the Hessian matrix 
  $\frac{\partial^2 H}{\partial I_l \partial I_l'}$
  at any point is non-negative. 
\item[(A4)] Each action takes positive values $I_l\ge 0$ and is not
  bounded from above.  This assumption excludes that any of the
  variables $q^l$ in which the wave function separates is cyclic. This
  is less restrictive than it may appear: for a system with rotational
  invariance one may reduce the attention either to functions which
  are even or odd under a reflection with respect to a hyperplane
  through the axis of rotation.
\item[(A5)] We assume that the Hamilton function is a homogeneous
  function of the actions. For $\lambda>0$ we then have
  \begin{equation}
    H(\lambda \mathbf{I})= \lambda^\alpha H(\mathbf{I})
  \end{equation}
  where $\alpha>0$ is the degree of homogeneity.  This assumption
  implies that the dynamics in each energy shell is equivalent up to a
  scaling factor.  For free motion on a manifold one has
  $\alpha=2$, so  this assumption is mainly a restriction 
  on the potentials.
  Note that the harmonic oscillator in any dimension has degree
  $\alpha=1$.
\end{itemize}
The above assumptions are not completely independent.  Some may be
relaxed without distorting our discussion too much
(see Section \ref{discussion}).\\
The EBK spectrum of semiclassical energy eigenvalues is given by
\begin{equation}
  E^{\mathrm{EBK}}_{\mathbf{n}}= H(I_1=n_1+\mu_1,\dots,I_s=n_s+\mu_s)
\end{equation}
where the quantum numbers $n_l=0,1,2,\dots$ are non-negative integers,
and the shifts $\mu_l$ are fixed numbers of order unity.
E.g. the $s$-dimensional harmonic oscillator has $\mu_l=1/2$
for all $l$ and free motion on an $s$-dimensional cuboid with
Dirichlet boundary conditions has $\mu_l=1$.
For our discussion the actual value of $\mu_l$ is not relevant.\\
For a given set of quantum numbers the wave function can be written as
\begin{equation}
  \Phi_\mathbf{n}(\mathbf{q})= \prod_{l=1}^s
  \phi^{(l)}_{\mathbf{n}}(q^l)
\end{equation}
with real functions $\phi^{(l)}_{\mathbf{n}}(q^l)$ of one variable.
The corresponding nodal pattern will then have a checker board structure.  Each
of these functions obeys
Sturm's oscillation theorem, i.e.  $\phi^{(l)}_{\mathbf{n}}(q^l)$ contains
$n_l$ nodal points. For the explicit EBK wave functions this is 
straight forward to show. This implies that the number of nodal domains in
the wave function $\Phi_\mathbf{n}(\mathbf{q})$ is equal to
\begin{equation}
  \nu_{\mathbf{n}}=\prod_{l=1}^s (n_l+1)\ .
\end{equation}
Note that for a degenerate spectrum separability implies a
definite and natural choice of preferred basis functions.

\subsection{The normalised nodal counts and Weyl's law}

In order to find the normalised nodal count $\xi_N= \nu_N/N$ for a
given wave function with quantum numbers $\mathbf{n}=(n_1,\dots,n_s)$
we need to know the spectral counting index $N\equiv
N(\mathcal{n})$.  An exact ordering of the quantum
numbers is a formidable task -- in the degenerate case one also needs to
make some choice for the order of basis functions with the same energy.  
In the present context any such order would be fine -- as it turns
out to leading order one only needs a sufficiently good approximation
to the exact counting index as provided by Weyl's law.
Indeed the semiclassical approximation
we use introduces an error in the ordering which may easily exceed
any influence of degeneracies.
Weyl's law states that
\begin{equation}
  N(E) \sim \mathcal{V}_\Gamma E^{s/\alpha}
  \label{Weylslaw}
\end{equation}
gives the leading asymptotic order of the spectral counting index as
$E \to \infty$. Here
\begin{equation} 
  \mathcal{V}_\Gamma=\int_\Gamma d\mathbf{I}
\end{equation} 
is the volume of the region $\Gamma\equiv\{\mathbf{I}:0 \leq
H(\mathbf{I})\leq 1\}$ in action space. For a free particle it is
related to the volume $\mathcal{V}_\mathcal{M}$ of the manifold by
$\mathcal{V}_\Gamma= \mathcal{V}_\mathcal{M}
\frac{\mathcal{V}_{B_s}}{(2\pi)^s}$ where $\mathcal{V}_{B_s}=
\frac{\pi^{s/2}}{\Gamma\left(\frac{s}{2}+1\right)}$ is the volume of
the $s$-dimensional unit ball.

Weyl's law (\ref{Weylslaw}) allows us to write the asymptotic
expression
\begin{equation}
  \xi_{\mathbf{n}}\sim \frac{\prod_{l=1}^s (n_l+1)}{\left[H(\mathbf{n}+
      \boldsymbol{\mu})\right]^{s/\alpha} 
    \mathcal{V}_\Gamma} \sim 
  \frac{\prod_{l=1}^s n_l}{E_{\mathbf{n}}^{s/\alpha} \mathcal{V}_\Gamma}
  + \mathcal{O}(E_{\mathbf{n}}^{-1/\alpha})
  \label{xi_asym}
\end{equation}
for the normalised nodal count. The error estimate on the right side of
(\ref{xi_asym}) is based on the homogeneity of the Hamilton function which 
implies $n_l \sim E^{1/\alpha}$. 
Expression (\ref{xi_asym}) will serve as the
starting point of the derivation of the limiting distribution in
section \ref{limdist}. 

Let us now derive Weyl's law (\ref{Weylslaw}) in the present setting.
This will not only serve us to estimate next-to leading orders in the
asymptotic formulas (\ref{Weylslaw}) and (\ref{xi_asym}) but also give
us an opportunity to introduce further details of the setting.  
The exact spectral counting function is defined by
\begin{equation}
  N(E)=\# \{ E_\mathbf{n} \le  E\}= \sum_{\mathbf{n}} \Theta(E-E_n)
\end{equation}
where $\Theta(x)$ is Heaviside's unit step function. Replacing the
exact energies by the EBK approximation introduces a small error by
shifting the positions of the steps slightly. The error introduced by
this shift is much smaller than the fluctuations in the spectral
counting function around its mean value and will  be neglected.  
The Poisson
summation formula in the form 
\begin{equation}
  \label{Poisson_sum}
  \sum_{n=0}^\infty F(n)=
  \sum_{M=-\infty}^\infty \int_{-1/2}^\infty e^{2\pi M x} F(x) dx
\end{equation} 
and
the homogeneity of the Hamilton function asserted by assumption (A5)
allow us to write 
\begin{eqnarray}
  N(E) =& E^{s/\alpha}
  \sum_{\mathbf{M}} \int_{I_l>\frac{\mu_l-1/2}{E^{1/\alpha}}} e^{2\pi
    i E^{1/\alpha}
    \mathbf{M}\cdot  \mathbf{I}- 2\pi i \mathbf{M}\cdot\mathbf{\mu}} \Theta(1-
  H(\mathbf{I})) d^s\mathbf{I} \label{Poisson}\\
  =& 
  \overline{N}(E) + N_{\mathrm{osc}}(E)
  \label{meanosc}
\end{eqnarray}
where
\begin{equation}
  \overline{N}(E)= E^{s/\alpha}\left( \int_{\Gamma} d^s\mathbf{I}
    +\mathcal{O}(E^{-1/\alpha}) \right) \sim E^{s/\alpha} V_\Gamma
\end{equation} 
is the contribution from the non-oscillating integral
$M_1=M_2=\dots=M_s=0$ and $N_{\mathrm{osc}}(E)$ is the sum over all
remaining (oscillating) integrals -- each being at most of order
$E^{(s-1)/\alpha}$.  Altogether we have derived Weyl's law
(\ref{Weylslaw}) and estimated that the sub-leading correction is 
a factor of order $E^{1/\alpha}$ smaller than the leading term.

\subsection{The geometry of the energy shell and rescaled actions}\label{sec_geom}

\begin{figure}[h]
  \centering
  \includegraphics[width=0.9\textwidth]{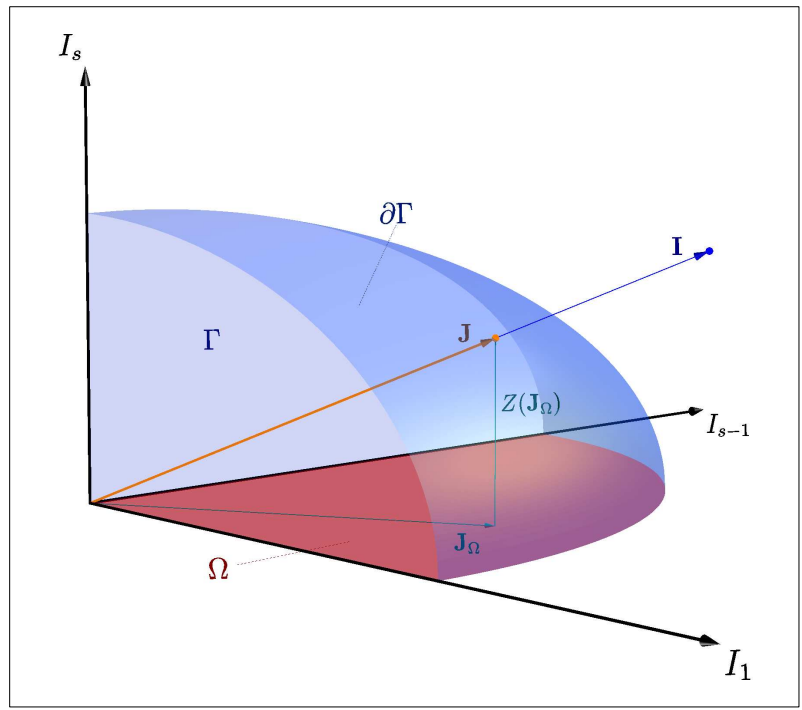}
  \caption{Illustration of  the region
    $\Gamma=\{\mathbf{I}:H(\mathbf{I})\le 1\}$ in
    $s$-dimensional action
    space. 
    The $s-1$-dimensional hyperplane $I_s=0$ is represented by a two-dimensional 
    plane spanned by the $I_1$ and $I_{{s-1}}$ axes in this picture.
    The illustration also shows the hyper-surfaces 
    $\partial \Gamma$ (the unit energy shell, i.e. the level set of the Hamilton function
    $H(\mathbf{I})=1$) and  $\Omega$.
    These  are the upper and lower
    parts of the boundary of $\Gamma$.\\
    A general point in action space with coordinates $\mathbf{I}$ is
    projected onto
    the  unit energy shell $\partial \Gamma$ where it is represented
    by the $s$-tuple $\mathbf{J}$.
  }
  \label{figgammageom}
\end{figure}

\begin{figure}[h]
  \centering
  \includegraphics[width=0.9\textwidth]{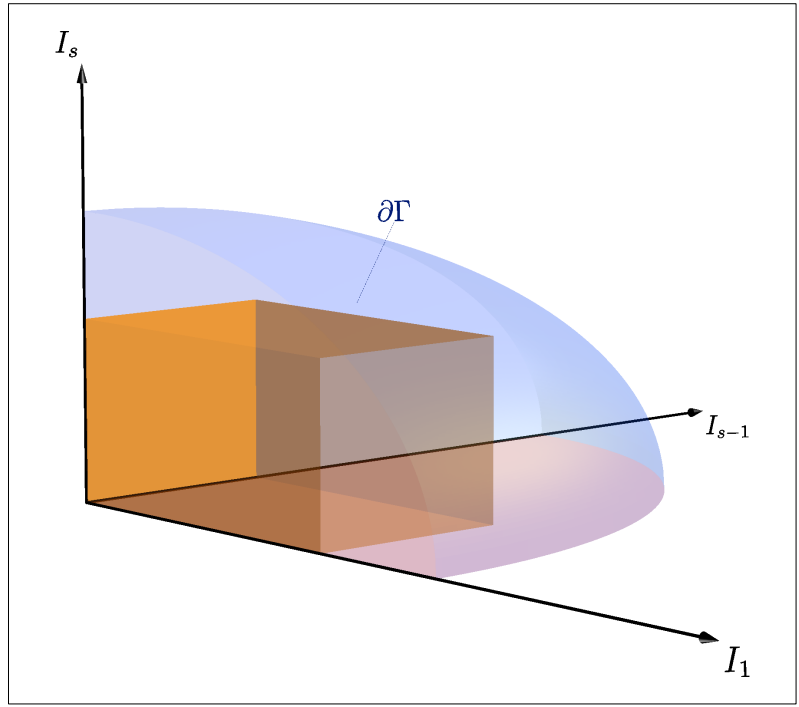}
  \caption{Illustration of  the geometric interpretation of the
    normalised nodal count as a ratio of two volumes
    $\xi(\mathbf{J}_\Omega)=\mathcal{V}(\mathbf{J}_\Omega)/\mathcal{V}_\Gamma < 1$.
    The cuboid of volume $\mathcal{V}(\mathbf{J}_\Omega)$ 
    is inscribed in the region
    $\Gamma$ with volume $\mathcal{V}_\Gamma$. The faces of the cuboid
    are parallel to the hyper-surfaces $I_l=0$ with one corner at the
    origin and the opposite corner on the surface $\partial \Gamma$.
  }
  \label{figratio}
\end{figure}

It is worth looking at the geometry of the region $\Gamma$
and the hyper-surface $\partial\Gamma$ in more
detail (see  Figure \ref{figgammageom} for an illustration).  
Indeed we here deal integrals of the type
\begin{equation}
  \mathcal{I}= \int_{E\cdot \Gamma} f(\mathbf{I}) d^s I
  \label{exint}
\end{equation}
with a homogeneous function $f(\mathbf{I})$ (of order $\beta$).  
Here $E\cdot\Gamma=\{ \mathbf{I}: H(\mathbf{I})\le E \}$ is a 
scaled version of the region $\Gamma$. 
Note that $\Gamma$ is compact and convex
--  compactness follows from assumption (A1)
and convexity from the second part of assumption (A3).
Indeed, compactness of the classically allowed region
$\mathcal{V}_E$ implies compactness of the region
$\mathcal W=\{(\mathbf p,\mathbf q):H(\mathbf p,\mathbf q)\le 1\}$
in phase space because the allowed momenta for any point $\mathbf{q}\in 
\mathcal{V}(E)$
form a closed $s$-dim ball in the cotangent space $T^* \mathcal
M_{\mathbf{q}}$. 
Describing the region $\mathcal{W}$ in action-angle
variables eventually implies compactness.\\
Using homogeneity one may reduce the 
$s$-dimensional integral (\ref{exint}) 
to an $s-1$-dimensional
integral over the $s-1$-dimensional compact surface (unit energy shell)
$\partial\Gamma\equiv \{\mathbf{I}:H(\mathbf{I})=1\}$ in momentum
space. Note that $\partial \Gamma$ is the non-trivial part of the
boundary of the region $\Gamma$ and
intersects the hyperplanes $I_l=0$ (which are also boundaries of
$\Gamma$). \\ 
This reduction is performed by a substitution
to rescaled action variables. The latter are defined by
\begin{equation}
  \mathbf{I}= \varepsilon \mathbf{J}
  \label{scaledaction}
\end{equation}
such that
\begin{equation}
  H(\mathbf{J})=1.
  \label{Jvar}
\end{equation} 
The rescaled action variables $J_l$ are not independent. Assumption
(A3) allows us to use the implicit function theorem and solve
(\ref{Jvar}) for
\begin{equation}
  \hat{I}_s=Z_{\Gamma}(J_1,\dots,J_{s-1})\ .
\end{equation}
We will denote the $s-1$-tuple of rescaled actions that appears as the
argument by $\mathbf{J}_\Omega\equiv(J_1,\dots,J_{s-1})$ such that
$\mathbf{J}= (\mathbf{J}_\Omega, Z_\Gamma(\mathbf{J}_\Omega))$.  Again using
assumption (A2) one can show that the function
$Z_{\Gamma}(\mathbf{J}_\Omega)$ is a decreasing of all arguments because
\begin{equation}
  \frac{\partial Z_\Gamma}{\partial J_l} =
  -\frac{\omega_l(\mathbf{J}_\Omega,Z_\Gamma(\mathbf{J}_\Omega))}{\omega_s(\mathbf{J}_\Omega,
    Z_\Gamma(\mathbf{J}_\Omega))
  }
\end{equation}
by the implicit function theorem.  The intersection of $\partial
\Gamma$ with $J_s=Z_\Gamma(\mathbf{J}_\Omega)=0$
marks the boundary of the range $\Omega$ of the variables $\mathbf{J}_\Omega$.\\
Let us now come back to the transformation (\ref{scaledaction}). 
It implies a change of integration variables
from the $s$ unscaled actions $\mathbf{I}$ to the independent
$s-1$ scaled actions
$\mathbf{J}_\Omega$ with values in the region $\Omega$ and a scaling factor
$\varepsilon \in [0,E]$.  The Jacobean can be calculated straight forwardly and
is given by
\begin{equation}
  \mathcal{J}= \varepsilon^{s-1}\left(
    Z_{\Gamma}(\mathbf{J}_\Omega) -\mathbf{J}_\Omega\cdot \nabla_{\mathbf{J}_\Omega}
    Z_\Gamma(\mathbf{J}_\Omega)\right)
  = \varepsilon^{s-1}\frac{\alpha}{\omega_s(\mathbf{J})}
\end{equation}
where the right hand-side follows from Euler's homogeneous function
theorem for the Hamilton function. The Jacobean is thus positive for
$\mathbf{J}_\Omega\in \Omega$ and $\varepsilon\in (0,E]$.  With the
shorthand
\begin{equation}
  d\Gamma=  \left(
    Z_{\Gamma}(\mathbf{J}_\Omega) -\mathbf{J}_\Omega\cdot \nabla_{\mathbf{J}_\Omega}
    Z_\Gamma(\mathbf{J}_\Omega)\right) \prod_{l=1}^{s-1} dJ_l
\end{equation}
we may now rewrite (\ref{exint}) as
\begin{equation}
  \label{reduced_integral}
  \mathcal{I}=
  \frac{E^{s+\beta}}{s+\beta}
  \int_\Omega d\Gamma f(\mathbf{J}_\Omega, Z(\mathbf{J}_\Omega)) \ .
\end{equation}
For $f\equiv 1$ this implies
\begin{equation}
  \mathcal{V}_\Gamma= \frac{1}{s} \int_\Omega d \Gamma .
\end{equation}
It is worth giving a geometrical illustration of the 
asymptotic nodal count (\ref{xi_asym}). In rescaled action variables
$n_l\equiv I_l= \varepsilon J_l$ the normalised nodal counts becomes a ratio
\begin{equation}
  \xi_{\mathbf{n}}\sim
  \frac{\mathcal{V}(\mathbf{J}_\Omega)}{\mathcal{V}_\Gamma}
  \label{xi_ratio}
\end{equation} 
where
\begin{equation}
  \mathcal{V}(\mathbf{J}_\Omega)= \left(\prod_{l=1}^{s-1} J_l\right) Z(\mathbf{J}_\Omega)
\end{equation}
is the volume of an $s$-dimensional cuboid in action space with faces parallel to
the hyperplanes $I_l=0$, and with one corner in the origin and the other on
a point $\mathbf{J}=(\mathbf{J}_\Omega,Z(\mathbf{J}_\Omega))$ on the surface $\partial \Gamma$
(see figure \ref{figratio} for an illustration).
As $\mathcal{V}(\mathbf{J}_\Omega) < \mathcal{V}_\Gamma$ we immediately obtain
$\xi(\mathbf{J}_\Omega)<1$ which is consistent with 
Courant's theorem \cite{Courant}. Since the maximal value of the volume
$\mathcal{V}(\mathbf{J}_\Omega)$ is definitely smaller than $\mathcal{V}_\Gamma$
the result is also consistent with 
Pleijel's theorem \cite{Pleijel}. 
Let $\mathbf{J}_\mathrm{crit}$ be the values for the rescaled action
where $\mathcal{V}(\mathcal{J}_\Omega)$ takes its maximal value for $\mathbf{J}_\Omega\in
\Omega$. Then $\mathbf{J}_\mathrm{crit}$ is a solution of the 
equations
\begin{equation}
  Z(\mathbf{J}_\Omega) = - J_l \frac{\partial Z (\mathbf{J}_\Omega)}{\partial J_l}
  \qquad l=1,\dots,s-1 \ .
  \label{crit_Vol}
\end{equation} 
Note that the left hand side is a strictly decreasing function
of $J_l$ while
convexity of $\Gamma$ implies that the right 
hand side is an increasing function.
As a consequence the solution to equation (\ref{crit_Vol}) is unique and
$\mathcal{V}(\mathbf{J}_\Omega)$ has only one critical point
in $\Omega$ which is the global maximum.\\
In the asymptotic regime $E \to \infty$
there will be no normalised nodal counts which are 
larger than the critical value
\begin{equation}
  \xi_\mathrm{crit} = 
  \frac{\mathcal{V}(\mathbf{J}_\mathrm{crit})}{\mathcal{V}_\Gamma}< 1\ .
\end{equation} 
Simple geometric intuition based on this picture shows that
$\xi_\mathrm{crit}$ will usually not be very close to either zero or unity 
for moderate dimensions -- in high dimensions one may have
$\xi_\mathrm{crit}\ll 1$.

\section{The nodal count distribution and its universal properties}
\label{limdist}

Let us now consider the nodal domain distribution (\ref{ndomdistr}).
Poisson summation (\ref{Poisson_sum}) and an application
of (\ref{reduced_integral})  then gives
$P(\xi)_{I_g(E)} \sim P(\xi) +\mathcal{O}(E^{-1/\alpha})$ with the
limiting distribution
\begin{equation}
  P(\xi)= \frac{1}{s \mathcal{V}_\Gamma}\int_\Omega d\Gamma\
  \delta\left(\xi-\frac{ \mathcal{V}( \mathbf{J}_\Omega ) }{\mathcal{V}_\Gamma}\right)\ .
  \label{Pxi_lim}
\end{equation}
The limiting distribution above does not depend on the size $g$
of the spectral interval $ I_g(E)=[E,(1+g)E]$. 
Note also that $P(\xi)$ is obtained as a weak limit, i.e. in the sense of weak
convergence of linear functionals which (together with the fact that
the support is always finite) ensures convergence of all
moments.\\
In practice one may consider
$P(\xi)_{I_g(E)}$ numerically in the form of a histogram
(i.e. in a locally averaged form)
and these will have some corrections to the limiting distribution --
these corrections will depend on the energy $E$, the size $g$ of the
spectral interval, and the bin size that has been used for the histogram.
As $E\to \infty$ with $g$ and bin size fixed
the fluctuation will become smaller. Indeed one may decrease the bin size
moderately 
as $E$ increases -- for convergence to a smooth function one just has to ensure
that the number of normalised nodal counts per bin increases indefinitely.
 
Expression (\ref{Pxi_lim}) is quite general and we will now turn
deriving some 
universal properties by a close analysis of this expression.
In section \ref{sec_geom} we have already mentioned that there
is an upper bound $\xi_\mathrm{crit}<1$ to the normalised nodal count.
This implies a cut-off for  the nodal domain distribution
$P(\xi)$ which has its support inside $0\le \xi \le \xi_\mathrm{crit}$.\\
Within its support $P(\xi)$ is differentiable.
This follows from the fact that  $V(\mathbf{J}_\Omega)$ has 
only
one critical point (maximum) in $\Omega$. 
At $\xi=0$ and $\xi=\xi_\mathrm{crit}$ the distribution $P(\xi)$ may have
singularities. We will show that the behaviour at the cut-off 
is mainly governed by the dimension $s$.
For $s=2$ one has a square root divergence, for $s=3$ there is a
finite step, 
and for $s\ge 4$
the distribution becomes continuous at $\xi=\xi_\mathrm{crit}$
but not smooth.

\subsection{The behaviour of $P(\xi)$ near the cut-off $\xi_\mathrm{crit}$.}

For $\xi$ smaller and close to $\xi_\mathrm{crit}$ the contribution to
$P(\xi)$ depend on the behaviour of $\mathcal{V}(\mathbf{J}_\Omega)$ near its
maximal value which it takes at $\mathbf{J}_\mathrm{crit}$.
Taylor expansion of $\mathcal{V}(\mathbf{J}_\Omega)$ to second order
around the maximum gives
\begin{equation}
  \frac{\mathcal{V}(\mathbf{J}_\Omega)}{\mathcal{V}_\Gamma}=
  \xi_\mathrm{crit} - \sum_{l,l'=1}^{s-1} \mathcal{H}_{ll'} \Delta J_l
  \Delta J_{l'} +\mathcal{O}(\Delta J^3)
\end{equation}
where $\Delta J_l=J_l-J_{\mathrm{crit},l}$ and $\mathcal{H}_{ll'}$ is
a positive definite matrix.
From
\begin{equation}
  d\Gamma = \left( s Z(\mathbf{J}_\mathrm{crit}) +\mathcal{O}(\Delta
    J)\right) \prod_{l=1}^s dJ_l
\end{equation}
one obtains
\begin{equation}
  P(\xi) \sim
  \frac{Z(\mathbf{J}_\mathrm{crit}) \mathcal{V}_{\mathcal{S}^{s-2}}}{2
    \mathcal{V}_\Gamma \sqrt{\det \mathcal{H}}}
\end{equation}
where $ \mathcal{V}_{\mathcal{S}^{s-2}} = 2
\pi^{(s-1)/2}/\Gamma((s-1)/2)$
is the volume of the $s-2$-dimensional sphere. 
The two dimensional case $s=2$  is included in this analysis, with
$\mathcal{V}_{\mathcal S^{0}}=2$. In this case
$P(\xi)$ diverges $\propto \frac{1}{\sqrt{\xi_\mathrm{crit} -\xi}}$
as was shown before in \cite{BGS}. When $s=3$ we observe that
$P(\xi) \to \mathrm{const}>0$ as $\xi \to \xi_\mathrm{crit}$ from below.
For $s\ge 4$
we have $P(\xi)\propto (\xi_\mathrm{crit}-\xi)^{(s-3)/2} \to 0$ such that $P(\xi)$
is continuous at $\xi=\xi_\mathrm{crit}$.

\subsection{The behaviour of $P(\xi)$ near $\xi = 0$.}

Now, we will study the behaviour of $P(\xi)$ near $\xi=0$. 
For $s=2$  it is not difficult to show
that
\begin{equation}
  \lim_{\xi\to 0^+ }P(\xi)=
  \frac{1}{2}\lim_{J_1\to 0} \frac{Z(J_1)-J_1 Z'(J_1)}{Z(J_1)+J_1
    Z'(J_1)}
  +
  \frac{1}{2}\lim_{Z \to 0} \frac{J_1(Z) -Z J_1'(Z) }{
    J_1(Z) + Z  J_1'(Z)}
  =1
\end{equation}
where $J_1(Z)$ is the inverse function of $Z(J_1)$.\\
For the rest of this section we keep our focus on $s\ge 3$. Note that  
the $\delta$-function $\delta(\xi-
\mathcal{V}(\mathbf{J}_\Omega)/\mathcal{V}_\Gamma)$ 
in expression (\ref{Pxi_lim}) 
for $\xi < \xi_\mathrm{crit}$
reduces the integral to an $s-2$-dimensional integral over the
level surfaces of $\mathcal{V}(\mathbf{J}_\Omega)$. These are closed 
deformations of an $s-2$-dimensional sphere. 
For our present purpose it is useful to write 
\numparts
  \begin{eqnarray}
    d\Gamma&=& s d\Gamma_1+d\Gamma_2\\
    d\Gamma_1&=&  Z(\mathbf{J}_\Omega) \prod_{l=1}^{s-1} dJ_l\\
    d\Gamma_2&=& -\frac{Z(\mathbf{J}_\Omega)}{\mathcal{V}(\mathbf{J}_\Omega)}
    \left[
      \mathbf{J}_\Omega \cdot \nabla_{\mathbf{J}_\Omega}
      \mathcal{V}(\mathbf{J}_\Omega)\right] \prod_{l=1}^{s-1} dJ_l
  \end{eqnarray}
\endnumparts
We will  show  below that $d\Gamma_2$ does not give a contribution to
the nodal count distribution which then reduces to
\begin{equation}
  P(\xi) = \frac{1}{\mathcal{V}_\Gamma}\int d\Gamma_1 \ \delta\left( 
    \xi- \frac{\mathcal{V}(\mathbf{J}_\Omega)}{\mathcal{V}_\Gamma}
  \right) = \int_{S_\xi}
  \frac{Z(\mathbf{J}_\Omega)}{|\nabla_{\mathbf{J}_\Omega}
\mathcal{V}(\mathbf{J}_\Omega)|} dS_\xi
  \label{Pxi_lim2}
\end{equation} 
where
$dS_\xi$ is the surface volume (area) element of the surface
\begin{equation}
  S_\xi=\{\mathbf{J}_\Omega:\mathcal{V}(\mathbf{J}_\Omega)=\xi \mathcal{V}_\Gamma\}.
\end{equation}
In order to show that the corresponding integral over $d\Gamma_2$
vanishes one may start with
\begin{equation}
  \int_{\Omega} d\Gamma_2\ \delta\left( 
    \xi- \frac{\mathcal{V}(\mathbf{J}_\Omega)}{\mathcal{V}_\Gamma}\right)=
  -\int_{S_\xi}
  \frac{\mathbf{J}_\Omega\cdot\mathbf{n}}{\prod_{l=1}^{s-1} J_l} dS_\xi 
\end{equation}
where $\mathbf{n}=\nabla_{\mathbf{J}_\Omega}
\mathcal{V}(\mathbf{J}_\Omega)/|\nabla_{\mathbf{J}_\Omega}
\mathcal{V}(\mathbf{J}_\Omega)|$
is the unit normal vector on the surface $S_\xi$. Gau{\ss}' theorem turns this
into a volume integral over the region
$\mathcal{V}(\mathbf{J}_\Omega)>\xi \mathcal{V}_\Gamma$ enclosed by
the surface. The corresponding integrand is the divergence
of the vector $\frac{\mathbf{J}_\Omega}{\prod_{l=1}^{s-1} J_l}$ which
vanishes identically which proves that equation (\ref{Pxi_lim2}) is correct.

\begin{figure}[h] 
  \centering 
  \includegraphics[width=0.9\textwidth]{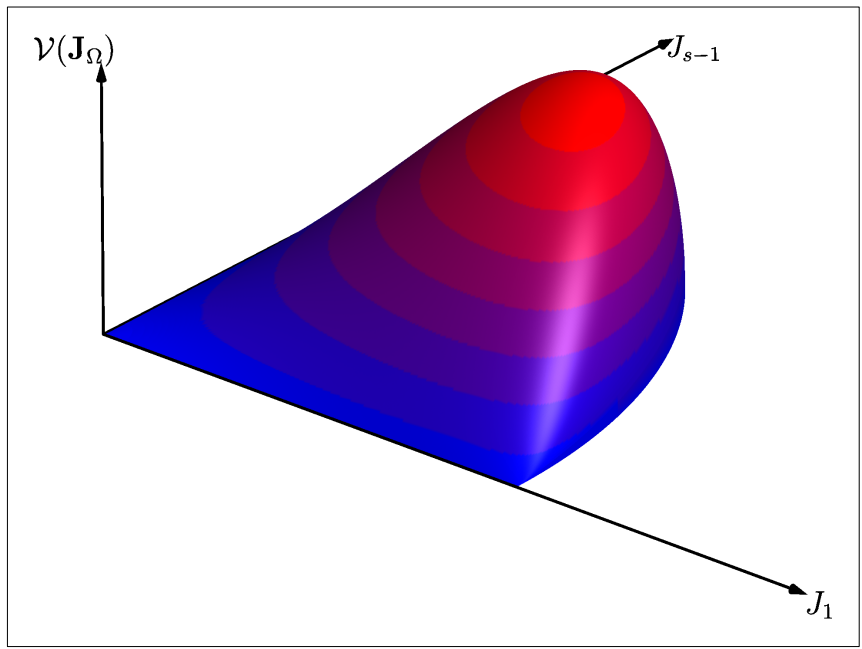}
  \caption{
    Illustration of the graph of the function
    $\mathcal{V}(\Gamma)$ over the region $\Omega$.
    For $s=3$ this illustration is exact and one can see three
    saddles in the corners of the region $\Omega$.
    Near these corners, especially near the one at the origin, the values of $\mathcal{V}(\Gamma)$ 
    are strongly suppressed.\\
    For $s>3$ one the illustration has to be taken with some care as it
    represents an
    $s-1$-dimensional plane by a two-dimensional -- the strong
    suppression is actually enhanced in this case.
  }
  \label{figcorners}
\end{figure}

For $\xi \to 0^+$ the surface 
$S_\xi$
approaches the boundary $\partial \Omega$ of $\Omega$. 
From (\ref{Pxi_lim2}) we see that the contributions from 
a volume element $dS_\xi$ carry a weight $Z(\mathbf{J}_\Omega)/
|\nabla_{\mathbf{J}_\Omega} \mathcal{V}(\mathbf{J}_\Omega)|$, so it will be 
dominated by any critical points 
where $|\nabla_{\mathbf{J}_\Omega} \mathcal{V}(\mathbf{J}_\Omega)|=0$
close to $S_\xi$.
Indeed there 
are such critical points on the boundary $\partial \Omega$
and they  
coincide with the set of points where $\partial \Omega$ is not 
smooth.
These
are the $s-3$-dimensional
intersections of any 2 hyperplanes $J_l=0$ or of one such hyperplane
with $Z(\mathbf{J}_\Omega)=0$. All of these are saddle points.\\
For $s=3$ 
the saddles are isolated at the three 
corners of $\partial \Omega$. The Hessian at these saddle points is
not degenerate.\\
For $s\ge 4$ the saddles are not isolated and the Hessian is degenerate. 
The saddles form continuous surfaces and where they intersect
the suppression
$|\nabla_{\mathbf{J}_\Omega} \mathcal{V}(\mathbf{J}_\Omega)|$ 
close to the intersection is enhanced  by the combined effect
of two or more intersecting saddle point surfaces.\\
The $s$ corners $\mathbf{J}^{(c)}$ ($c=0,\dots,s-1$)
of $\Omega$ thus dominate the $P(\xi)$ for $\xi\to 0^+$
(see figure \ref{figcorners} for an illustration). 
Explicitly the corners are given by
the origin $\mathbf{J}^{(0)}$ and the $s-1$ points 
$\mathbf{J}^{(c)}$
where
$Z(\mathbf{J}_\Omega)=0$ intersects with the $s-2$
hyperplanes of the form $J_l=0$ where $l\in \{1,\dots,s-1\}- \{c\}$.
At the corners the lowest order of a derivative which does not vanish
identically is $(s-1)$. 
We will show that this leads to a divergence
of the nodal count distribution $\lim_{\xi\rightarrow
  0^+}P(\xi)=\infty$.\\
We may focus on the 
leading contribution from the corner at the origin
which dominates the distribution for small
$\xi$ -- indeed the weight $Z(\mathbf{J}_\Omega)/
|\nabla_{\mathbf{J}_\Omega} \mathcal{V}(\mathbf{J}_\Omega)|$
suppresses the contribution at the corners due to the factor $Z(\mathbf{J}_\Omega)$
which is zero for all corners apart from the origin.\\
In order to derive the contribution from the origin let us start
by expanding enumerator and denominator
of the weight $Z(\mathbf{J}_\Omega)/
|\nabla_{\mathbf{J}_\Omega} \mathcal{V}(\mathbf{J}_\Omega)|$
independently. For the denominator one has
$|\nabla_{\mathbf{J}_\Omega} 
\mathcal{V}(
\mathbf{J}_\Omega)|=\mathcal{O}(\Delta \mathbf{J}_\Omega^{s-1})$. 
The enumerator $Z(\mathbf{J}_\Omega)$
however remains finite
$Z(\mathbf{J}_0)=Z_0>0$ near the origin.
Now consider the contribution 
\begin{equation}
  P(\xi) \sim  \frac{Z_0}{\mathcal{V}_\Gamma}
  \int_C 
  \delta\left( \xi - \frac{ Z_0 \prod_{l=1}^{s-1}J_l}{\mathcal{V}_\Gamma}\right)
  \prod_{l=1}^{s-1} dJ_l
  \label{limbeh0}
\end{equation}
from a small region $C$ that contains the origin.
 We have used $Z(\mathbf{J}_\Omega)\sim Z_0$  with corrections 
$\mathcal{O}(\mathbf{J}_\Omega)$.
The calculation is simplified if we take 
$C$ as an  $s-1$-dimensional cuboid
with side lengths $a_l$
\begin{equation}
  C = \{\mathbf{J}_\Omega: 0\le J_l \le a_l, l=1,\ldots,s-1 \} \ .
\end{equation}
The actual values of the side lengths $a_l$
will not enter the leading order which implies that we have a true corner
phenomenon and that the leading asymptotic order of integral does not depend on the details
of region $C$. \\
In order to perform the integration set
\begin{equation}
  \zeta=\frac{Z_0 \prod_{l=1}^{s-1} J_l}{\mathcal{V}_\Gamma}
\end{equation}
and transform coordinates
$\mathbf{J}_\Omega \rightarrow (J_1,\dots, J_{s-2}, \zeta)$.
The $\zeta$-integral can be performed and leaves
\begin{equation}
  P(\xi) \sim
  \int_0^{a_1} \frac{dJ_1}{J_1} \dots  \int_0^{a_{s-2}} \frac{dJ_{s-2}}{J_{s-2}}\
  \Theta\left( J_{s-2} -
    \frac{\mathcal{V}_\Gamma \xi}{a_{s-1} Z_0 \prod_{l=1}^{s-3} J_l }\right)
\end{equation}
where the factors $J_l^{-1}$ stem from the Jacobean.
This integral can be solved iteratively using
\begin{equation}
  \int_0^a \frac{dx}{x} (\log x)^l \Theta(x- b)
  =\Theta(a-b)\frac{(\log a)^{l+1} -(\log b)^{l+1}}{l+1}
\end{equation}
and gives the leading contribution
\begin{equation}
  P(\xi)\sim \frac{(- \log \xi)^{s-2}}{(s-2)!}\ .
  \label{Pxi_zero}
\end{equation}
Note that the leading order corrections to this depend on the
lengths $a_l$ which implies that a global approach is necessary to
evaluate the next-to-leading order of $P(\xi)$ as $\xi \to 0^+$.
We see that the leading order diverges 
as an $s-2$-th power of a logarithm
with a universal constant $1/(s-2)!$. Any system dependent
features can only enter at next-to-leading order.

\subsection{Monotonicity}

For $s=2$ one can check directly that $P(\xi)$ is a strictly
increasing function for $0< \xi< \xi_\mathrm{crit}$.
Indeed (\ref{Pxi_lim2}) is valid for $s=2$ and gives
\begin{equation}
  P(\xi) =
  \frac{Z(J_{1,-})}{Z(J_{1,-})+ Z'(J_{1,-}) J_{1,-}}
  -
  \frac{Z(J_{1,+})}{Z(J_{1,+})+ Z'(J_{1,+}) J_{1,+}}
  \label{Pxi_dim2}
\end{equation}
where $J_{1,-}<J_{1,+}$ are the two solutions of $\xi= J_1 Z(J_1)$.
Note that both terms in (\ref{Pxi_dim2}) are positive.
Differentiation of the first term yields
\begin{equation}
  \frac{J_{1,-}Z'(J_{1,-})^2-Z(J_{1,-})Z'(J_{1,-})- 
    J_{1,-}Z(J_{1,-})Z''(J_{1,-})}{(Z(J_{1,-})+ 
    Z'(J_{1,-}) J_{1,-})^2}\frac{dJ_{1,-}}{d \xi}>0
\end{equation}
because $\frac{d J_{1,-}}{d \xi}>0$.
Analogously the derivative of the second term gives a positive contribution 
because  $\frac{d J_{1,+}}{d \xi}<0$.

For $s\ge 3$ our calculations above imply that $P(\xi)$ is a
decreasing function in a neighbourhood of $\xi=0$. For $s\ge 4$
we have also shown that $P(\xi)$ is a decreasing function near
$\xi=\xi_\mathrm{crit}$ (for $s=3$ our results are consistent with a
decreasing function).
This suggests that $P(\xi)$ may be a decreasing function over its
full support $0< \xi < \xi_{\mathrm{crit}}$
for $s\ge
3$. Such a conjecture is supported by all example calculations that we
have performed -- however we have not been able to prove it.

\section{Two simple examples: the harmonic oscillator and the cuboid}
\label{examples}
 
Let us now illustrate our results with a few examples that allow for
more explicit treatment.

\subsection{The $s$-dimensional harmonic oscillator}
\begin{figure}[h]
  \centering
  \includegraphics[width=0.85\textwidth,height=0.70\textheight]{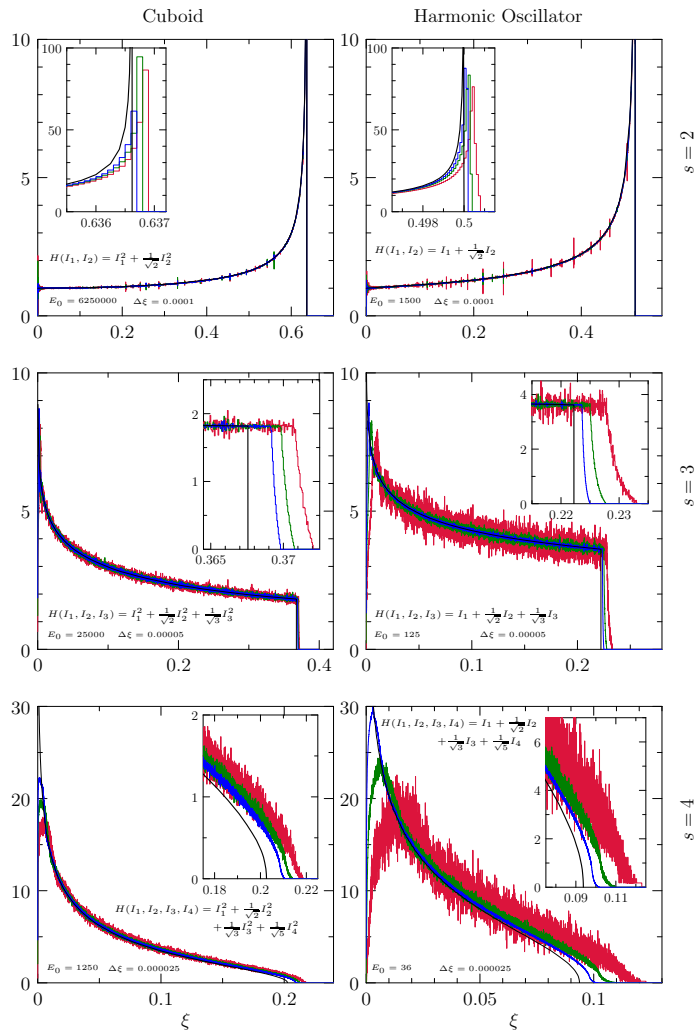}
  \caption{Nodal count distributions
    for the harmonic oscillator (right column) and the
    cuboid (left column) for $s=2$ (first row), $s=3$ (second row),
    and $s=4$ (third row). The black lines correspond to the
    limiting distributions $P_s(\xi)$. The red, green, and blue
    lines are numerically obtained histograms of the nodal count
    distribution
    at finite energy intervals: red line $[E_0,2E_0]$, green line
    $[2E_0, 4E_0]$, blue line $[4E_0,8E_0]$. The chosen values for $E_0$ are
    given in the corresponding graphs together with the chosen
    system parameters (the Hamiltonian), and the bin size $\Delta
    \xi$ that has been used for the histograms. 
    In each case normalised
    nodal
    counts have been obtained for the lowest 50 to 80 million eigenfunctions.\\
    The insets magnify the graph near the critical cut-off value
    $\xi_{\mathrm{crit}}$. Overall the numerically obtained
    histograms
    are consistent with the (weak) convergence to the limiting distribution.
  }
  \label{numerical_limit_dist}
\end{figure}

For a harmonic oscillator the Hamilton function is linear in the
action variables
\begin{equation}
  H(\mathbf{I}) = \sum_{l=1}^s \omega_l I_l
\end{equation}
The unit energy shell $\partial\Gamma$ in action space is then described by
the function
\begin{equation}
  J_s\equiv Z(\mathbf{J}_\Omega)=1 -\frac{1}{\omega_s}\sum_{l=1}^{s-1}
  \omega_l J_l\ .
\end{equation}
The volume of the region 
$\Gamma$ is $\mathcal{V}_\Gamma= \frac{1}{s!\prod_{l=1}^s \omega_l}$.\\
From $\mathcal{V}(\mathbf{J}_\Omega) = Z(\mathbf{J}_\Omega)
\prod_{l=1}^{s-1} J_l$ one finds its maximum value at $J_{\mathrm{crit},l}=\frac{1}{s \omega_l}$
such that $\mathcal{V}(\mathbf{J}_\mathrm{crit})= \frac{1}{s^s
  \prod_{l=1}^s \omega_l}$. This implies the critical value
\begin{equation}
  \xi_{\mathrm{crit},s}= \frac{s!}{s^s}
\end{equation}
for the normalised nodal count. Note that the individual frequencies
do not enter. Indeed 
the complete nodal count distribution with $s$
degrees of freedom does 
not depend
on the frequencies and can be expressed as 
\begin{eqnarray}
  P_s(\xi)= &s! \! \int_{\sum_{l=1}^{s-1} J_l \le1} \! \left(1-\sum_{l=1}^{s-1}
    J_l\right) \times \\
  & \quad \times
  \delta \!\left( \xi- s! \left(1-\sum_{l=1}^{s-1}
      J_l\right)\prod_{l=1}^{s-1} J_l \right) \prod_{l}
  dJ_l \ .
  \label{ho_Pxi}
\end{eqnarray}
For $s=2$ this integral has the explicit form
\begin{equation}
  P_2(\xi)=(1-2\xi)^{-1/2} \qquad \mathrm{for}\ \xi<1/2. 
\end{equation}
For arbitrary $s$ one may evaluate all positive integer moments
\begin{equation}
  \langle \xi^m \rangle_s\equiv\int_0^{\xi_{\mathrm{crit},s}} \xi^m P_s(\xi) d\xi = \frac{s!^{m+1} (m!)^s (m+1)}{(s(m+1))!}\ .
\end{equation}
See Figure \ref{numerical_limit_dist} for the graph of the limiting
distribution (\ref{ho_Pxi}) for $s=2,3,4$ together with numerical data
obtained for finite energy intervals.

\subsection{The $s$-dimensional cuboid}

We consider the free particle in an $s$-dimensional cuboid (rectangular box)
with side lengths $a_l$ ($l=1,\dots,s$). 
With Dirichlet conditions on the boundary of the box one obtains
exact energy eigenvalues
\begin{equation}
  E_{\mathbf{n}}=
  \pi^2\sum_{l=1}^s \frac{n_l^2}{a_l^2}
\end{equation}
where the quantum numbers $n_l$ run over positive integers.
The corresponding classical Hamilton function
$H(\mathbf{I})= \pi^2 \sum_{l=1}^s \frac{I_l^2}{s_l^2}$ is homogeneous
of order $\alpha=2$. The unit energy shell $\partial\Gamma$ is given in terms
of the function
\begin{equation}
  J_s\equiv Z(\mathbf{J}_\Omega)= \frac{a_s}{\pi}\sqrt{1-\pi^2 \sum_{l=1}^{s-1}
    \frac{J_l^2}{a_l^2}}\ .
\end{equation}
The volume of the region $\Gamma$ is
$\mathcal{V}_\Gamma=\frac{1}{\pi^{s/2} 2^{s-1} s \Gamma(s/2)
}\prod_{l=1}^s a_l$.\\
The maximal volume
$\mathcal{V}(\mathbf{J}_\Omega)$ of a cube touching the unit energy shell
is given by $\mathcal{V}(\mathbf{J}_{\mathrm{crit}})=\frac{1}{\pi^s
    s^{s/2}} \prod_{l=1}^s a_l$ (where $J_{\mathrm{crit},l}=
  \frac{a_l}{s^{1/2} \pi}$). 
One thus finds the critical value
\begin{equation}
  \xi_{\mathrm{crit},s}= \frac{2^{s-1} s \Gamma(s/2)}{\pi^{s/2} s^{s/2}}
\end{equation}
above which the limiting nodal domain distribution vanishes.
Similarly to the harmonic oscillator the critical value and the
limiting
distribution do
not depend on the detailed system parameters such as the side
lengths. 
Indeed the limiting distribution may be written as
\begin{eqnarray}
  P_s(\xi)=&\frac{2^{s-1}s\Gamma(s/2)}{\pi^{s/2}}
  \int_{\sum_{l=1}^{s-1} J_l^2 <1} \left(1-\sum_{l=1}^{s-1}
    J_l^2\right)^{1/2} \times\nonumber \\
  &\delta\left(\xi - \frac{2^{s-1}s\Gamma(s/2)}{\pi^{s/2}}
  \left(1-\sum_{l=1}^{s-1} J_l^2\right)^{1/2}\prod_{l=1}^{s-1} J_l \right)
  \prod_{l=1}^{s-1} dJ_l\ .
\end{eqnarray}
For $s=2$ this reduces to
\begin{equation}
  P_2(\xi)=\left(1-\frac{\pi^2 \xi^2}{4} \right)^{-1/2} \qquad
  \mathrm{for}\ \xi<\frac{2}{\pi}\ .
\end{equation}
See Figure \ref{numerical_limit_dist} for graphs of the distribution
for $s= 2, 3, 4$ together with numerically obtained histograms for
finite energy intervals.

\section{Discussion}
\label{discussion}

We have derived an expression for the limiting nodal count
distribution in the case where the wave equation is separable, and we
have
extracted some universal features of this distribution.
While we have formally limited the scope
with the assumptions (A1) to (A5) 
many standard examples as the harmonic
oscillator
or the particle in a cubic box obey all of these conditions.
For some other examples which do
not obey
all of the conditions it is straight forward to generalize our
derivations. For instance a particle in a spherical box does not obey
assumption (A4) as action variables that correspond to angular momenta
are not bounded from below. In this case the wave function separates
in variables, some of which are cyclic. The derivation of a limiting
distribution
follows in full analogy once the expression for the nodal count in
terms of quantum numbers is adapted and the Poisson summation is
performed
over the corresponding set of quantised action variables. 
Indeed most of our assumptions are purely technical and can be relaxed
if necessary -- though relaxing condition (A3) may imply that there
are additional local maxima of the volume
$\mathcal{V}(\mathbf{J}_\Omega)$ which may lead to further
singularities within the support of the limiting  nodal domain
distribution.
Also assumption (A5) that the Hamiltonian is a homogeneous function
can be relaxed to a certain degree. Indeed it is only needed that the energy shell at high
energies can be described asymptotically by a homogeneous function.

It would certainly be interesting to compare our results to nodal
count distributions of non-separable or wave-chaotic systems in
dimensions larger than two.
In any dimension one may try to obtain nodal counts numerically by using
a corresponding adaptation of the Hoshen-Kopelman algorithm \cite{hoshen}
and apply it to numerically obtained eigenfunctions. Berry's
conjecture \cite{berry-rw} states that  highly excited chaotic
eigenfunctions can be simulated by a Gaussian random wave
ensemble. In this ensemble the wave function is given by
\begin{equation}\label{bil rwm}
  \Phi_{\mathrm{RWM}}(\mathbf{q})=\mathrm{Re} \sqrt{\frac{2}{N}}\sum_{j=1}^N e^{ik\mathbf{n_j}\cdot \,\mathbf q + i\phi_j}
\end{equation}
where $\mathbf{n_j}$ are uniformly distributed on a unit
$(s-1)$-sphere and the phases $\phi_j$ are equidistributed on $[0,2\pi)$.
On dimensional grounds one expects that the number of nodal domain
in a given region of the random wave is proportional to
the volume of the region. For two-dimensional random waves this
has been checked and it is consistent with the critical percolation
conjecture \cite{Bogomolny}.
We tried to check this in three dimensions by finding the number
of nodal
domains of random waves inside a cube of side length $a$ (at fixed
wave number $k=1$). The artificial boundary of the cube leads
to nodal domains which intersect the boundary -- indeed we
have found that all nodal domains in our numerical
approach were intersecting the boundary and that
the nodal count is proportional to $a^2$ rather than $a^3$. This
scaling is expected on dimensional grounds for the number of nodal
domains
which intersect the boundary. 
However we were not able to
increase the side length beyond $a=100$ (about 16 wave lengths)
on a standard desktop and we have just looked at a few hundred realisations.
We cannot exclude that interior nodal domains
(those which do not touch the artificial boundary) start to appear in 
much larger cubes and eventually dominate the nodal count.
We can say however that any crossover from $a^2$ (boundary dominated)
to $a^3$ (bulk dominated) would have to occur at considerably higher
side lengths for which applying our numerical algorithm is
beyond the power of standard desktop
computers.\\
Our numerical findings do confirm the basic expectation
that the universality of
a critical percolation model does not apply in the three dimensional
case. For instance we find that
the volume of the largest nodal domain scales linearly
with the volume of the cube -- a clear indication that one is 
inside the (non-universal) percolating regime 
(a non-trivial exponent is expected at
the percolation transition).
We hope that future research will shed more light on the nodal sets
and nodal counts of wave-chaotic systems in dimensions $s\ge 3$ 
as well as in the corresponding random-wave models.

\end{document}